\documentclass[journal=jacsat,manuscript=article]{achemso}

\usepackage[version=3]{mhchem} 

\author{Itir Bakis Dogru Yuksel}
\author{Zhu Zhang}
\author{Marnix Vreugdenhil}
\author{Allard P. Mosk}
\author{Dries van Oosten}
\author{Sanli Faez}

\email{s.faez@uu.nl}
\affiliation[Utrecht University]
{Nanophotonics, Debye Institute for Nanomaterials Science, Utrecht University, Princetonplein 1, 3584 CC, Utrecht, The Netherlands}

\title[An \textsf{achemso} demo]
 {Laser-patterned Thin-film Electrodes: Imaging Ion Accumulation and Trapped Nanoparticles }
\abbreviations{IR,NMR,UV}

\begin{document}







\begin{abstract}

This study introduces a straightforward electrode design featuring sharp edges with a curvature of a few hundred nanometers in radius, with which both ion accumulation and nanoparticle deposition can be observed under an alternating electrical potential. The electrodes, termed 'shark-teeth electrodes', are fabricated using a laser ablation technique optimized for facile nanostructure creation. This method involves successive, overlapping ablated discs in a thin film of gold, producing sharp tips that generate strong electric fields. When electrically polarized in an electrolyte solution, these sharp tips form a screening layer, facilitating the observation of ion and nanoparticle behavior.

A total-internal reflection microscope is employed to monitor ion accumulation on these electrodes, demonstrating their capability in iontronic microscopy. Additionally, the same electrodes are used to track nanoparticle trapping under high-frequency alternating potentials. This dual functionality allows for the investigation of electrochemical and physical interactions between ions and colloidal nanoparticles, contributing valuable insights to the field of soft matter. 


\end{abstract}

\section{Introduction}
The manipulation and real-time observation of solvated ions and colloidal nanoparticles in liquids have applications in multiple disciplines, from environmental protection to technological innovation and life sciences. It aids in removing pollutants, contributing to cleaner ecosystems~\cite{zhu2020optimization}. This also allows for the alteration of material properties like electrical conductivity~\cite{rivnay2016structural}, thermal behavior~\cite{sharma2011enhancement}, and optical characteristics~\cite{lee2012improvement}, thereby driving advancements in materials science. In biophysics, controlling ion and particle movement is pivotal~\cite{warmlander2013biophysical,faez2015fast}. It drives research on molecular biology~\cite{lipovsek2014tracking}, yielding insights into cell behavior~\cite{foley2021mass} and innovation like lab-on-a-chip devices~\cite{hajjoul2009lab} for advancing biomedical investigations~\cite{cole2017label,young2018quantitative,young2019interferometric}.

A common method for manipulating colloidal species involves using an electric field, where the electrode design is critical for controlling the dynamics of ions~\cite{zhu2022insights,arnot2022thick} and particles~\cite{yantzi2007multiphase,dogruyuksel2024origami}. Recent studies have highlighted the importance of electrode design for innovative biomedical applications. Sy et al. proposed an electrokinetic chip utilizing a ring-shaped interdigitated electrode for rapidly collecting particles for fast bacterial counting~\cite{sy2023label}. Zhou et al. introduced a template electrokinetic assembly to showcase microparticle assembly at specific positions~\cite{zhou2020guided}. Ji et al. created a 3D model to analyze particle positions for cell manipulation~\cite{ji2020three}. Additionally, Varmazyari et al. developed a microfluidic system for cancer cell separation~\cite{varmazyari2022dielectrophoresis}. 

In addition to electrode design, optical microscopy techniques enable detailed observation of ion redistribution in response to an electric field at the liquid-solid interface, creating a potentiodynamic optical contrast. Ion currents are inferred from total-internal-reflection (TIR) microscopy~\cite{namink2020electric, zhang2023iontronic}, and interferometric reflection microscopy~\cite{utterback2023operando}, across various electrode configurations, where image analysis reveals ion accumulation. However, previous studies often focused solely on either ion concentration changes or particle locations.
In this study, we introduce a new electrode design that facilitates the simultaneous monitoring of both movements of ions and nanoparticles. Our planar electrodes are fabricated via an optimized laser ablation method, producing symmetrical sharp-patterned gold electrodes, referred to as 'shark-teeth electrodes'. This laser ablation technique is a facile method for creating nanostructures, involving successive ablated circles in a thin gold film to form sharp tips capable of generating strong electric fields.

We employed these 'shark-teeth electrodes' to analyze the potential frequency-dependent behavior of solvated species, ranging from ions to nanoparticles. By imaging the movement of ions and particles at various frequencies using TIR scattering and fluorescent microscopy, we provide detailed insights into the mechanisms governing their behavior. Our findings demonstrate that under an alternating current (AC) potential application at low frequencies (up to tens of Hz), both ions and nanoparticles oscillate due to electric double layer (EDL) modulation at the electrode nanoedges~\cite{namink2020electric}. However, at higher frequencies, the EDL does not have sufficient time to establish~\cite{saghafi2023high}, and the signal from ions diminishes. At these frequencies, the stronger electric field gradient formed at the sharp tips dominates nanoparticle drift motion and attracts them to these edges due to dielectrophoresis (DEP)~\cite{nili2016ac} effect.

Simultaneously monitoring ions and nanoparticles using shark-teeth electrodes holds promising implications for biological applications. These electrodes could be integrated into organs-on-a-chip devices~\cite{mastrangeli2019organ}, enabling real-time analysis of intricate biological processes. This advancement has the potential to significantly enhance clinical diagnostics~\cite{lu2015ac} and drug delivery~\cite{li2021design, ivanoff2016ac}.

\section{Results and discussion}
\subsection{Shark-teeth electrode fabrication}

\begin{figure}[!htbp] 
\centering
 \includegraphics[width=16 cm]{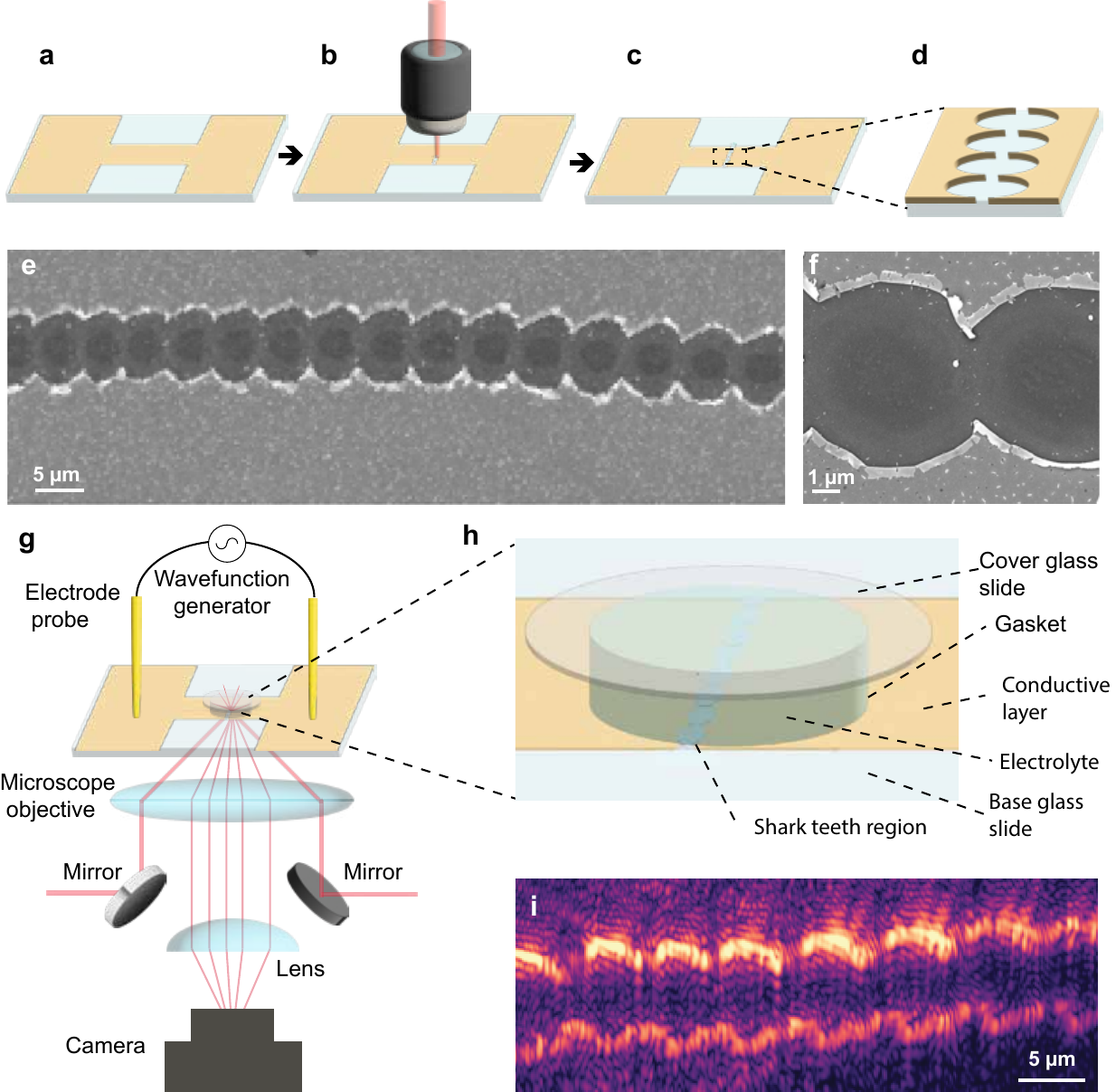}
  \caption{Shark-teeth electrode fabrication: (a) H-shaped conductive layer coated on a glass slide. (b) Laser ablation etching of the mid-region. (c) Resulting laser-ablated planar electrodes, and (d) with a zoomed version for detail. SEM images: (e) Shark-teeth electrodes, highlighting the periodic sharp edges on both electrodes; (f) Close-up view showing flaps at the corners of the laser-exposed regions. (g) Schematic of the ion flow analysis setup: ion flow is induced by a wavefunction generator through electrode probes on both of the shark-teeth electrodes and imaged by a TIR microscope. (h) Schematic of the shark-teeth electrodes depicting the electrolyte insertion between them within a gasket, covered by a glass slide.  (i) Visualization of the shark-teeth electrodes using TIR microscopy.}
  \label{fgr:electrode_fabrication}
\end{figure}

The planar thin-film electrodes were fabricated on a thin cover slide, featuring an H-shaped metallic coating composed of a chromium adhesion layer and a gold conductive layer (Figure~\ref{fgr:electrode_fabrication}-a). The metallic surface was then separated into two electrodes (Figure S1) using laser ablation, with successive femtosecond laser pulses focused onto the gold surface while it was moved horizontally through the laser focus (Figure~\ref{fgr:electrode_fabrication}-b); the setup for this process is detailed in Figure S2. This process resulted in the creation of features with sharp corners, resembling the structure of shark-teeth (Figure~\ref{fgr:electrode_fabrication}-c and d), hence we referred to them by that name. This design enhances electric field generation, particularly at the sharp tips. The periodicity of the sharp edges of symmetrical electrodes are shown in Figure~\ref{fgr:electrode_fabrication}-e while Figure~\ref{fgr:electrode_fabrication}-f highlights the details of the electrodes through Scanning Electron Microscopy (SEM) images. The ultrafast nature of laser pulses is crucial for inducing the observed folding at the edges of the gold layer, as depicted in Figure~\ref{fgr:electrode_fabrication}-f. Unlike longer laser pulses, femtosecond pulses do not lead to melting. Rather, the rapid energy deposition initiates an inertial heating process that leads to elastoplastic deformation of the gold film, causing the material to detach and form flaps~\cite{abram2024pre}.

The characterization of the developed electrodes, customized for investigating ion accumulation and trapping nanoparticles under potential modulation, involves the use of TIR microscopy,~\cite{zhang2021computing,zhang2023iontronic, peters2023dark} where the electric potential is generated via a wavefunction generator (Figure~\ref{fgr:electrode_fabrication}-g). Using TIR configuration instead of bright-field reflection helps by preventing the detection saturation in the camera and allows for optimally using the dynamic range of the camera for detecting the interference signal of light scattered from the ions around the nanoparticles and the (much) larger scattering from the nanoparticle, which here acts as a reference arm of an interferometer. We use a gasket to introduce the electrolyte and particle solutions between the electrodes and use a disk-shaped glass slide to cover the gasket and suppress solvent evaporation (Figure~\ref{fgr:electrode_fabrication}-h). Additionally, the visualization of the sample using TIR microscopy is presented in Figure~\ref{fgr:electrode_fabrication}-i, where the electrolyte solution is on top.



\subsection{Imaging ion concentration oscillations induced by the applied potential }
 In this section, we investigate the ion current between the gap of the shark-teeth electrodes in an aqueous solution with 0.1M potassium chloride (KCl) electrolyte when an AC potential is applied in the square waveform at 5 Hz and 1 V peak-to-peak. We intentionally applied low frequency to allow for the formation of the EDL and to observe changes in ion concentration. We recorded a video by TIR microscopy during the potential application to the shark-teeth electrode with KCl electrolyte solution on and between the electrodes. 
 Then the ion accumulation information was extracted from the series of frames (See Supplementary Video~1 for the periodic twinkling from the electrodes). 
 
\begin{figure}[!htbp] 
\centering
 \includegraphics[width=16 cm]{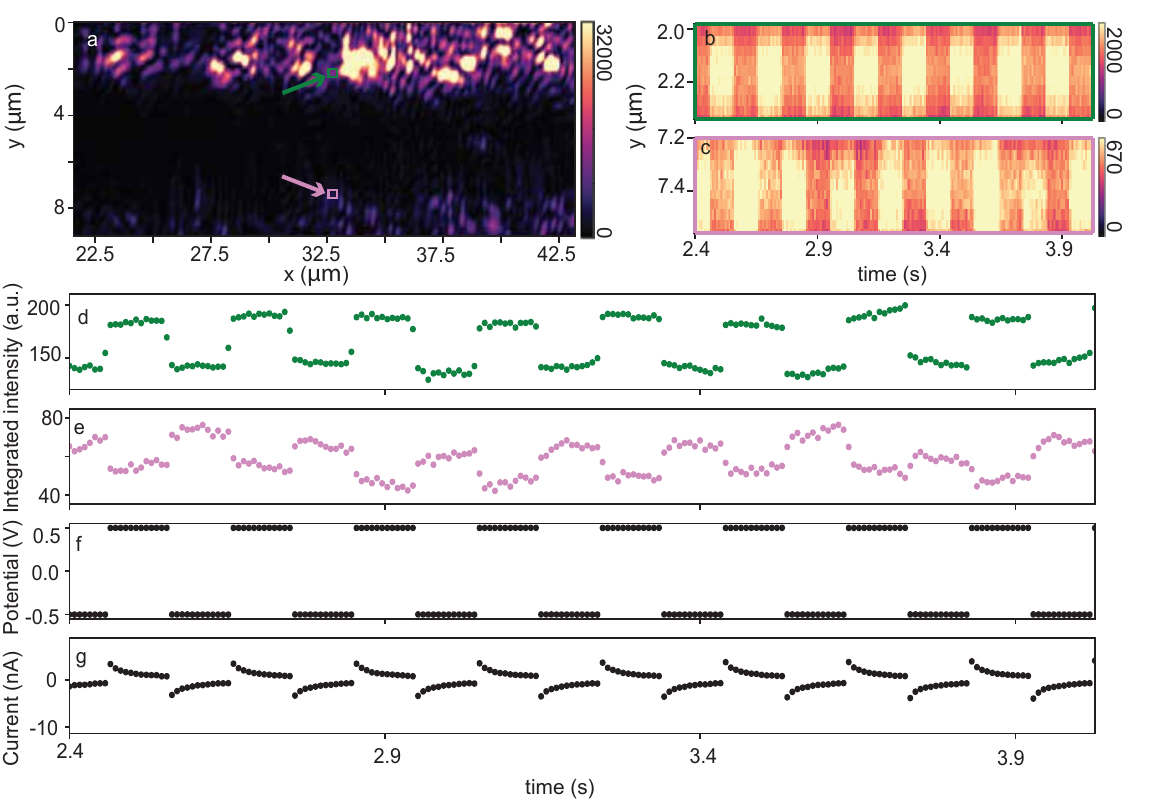}
  \caption{Real-time analysis of ion flow with scattering via shark-teeth electrode: (a) The shark-teeth electrode image by scattering under the influence of a 5 Hz, 1 V peak-to-peak square waveform. Two distinct regions marked with green and violet squares and arrows pointing them on opposite sides of the shark-teeth are selected for further examination. (b) and (c) Chemographs and, (d) and (e) integrated intensities with respect to time extracted from the green and violet marked regions, respectively. Corresponding plots of (f) applied electric potential and (g) measured current, confirm the in-operando ion concentration oscillations.}
  \label{fgr:imaging_field}
\end{figure}

Two square regions, each corresponding to 360 nm x 360 nm and marked in green and violet respectively (Figure~\ref{fgr:imaging_field}-a), were chosen for further examination, one on each of the opposite of the shark-teeth electrodes. The observed intensity changes, attributed to periodic EDL modulation synchronized with the applied potential, reflect the local ion concentration around the shark-teeth creating an optical contrast. In other words, we observe the potentiodynamic optical contrast resulting from the accumulation of ions during the reconfiguration of the EDL~\cite{namink2020electric} undergoing the potential modulation. This periodic contrast change enables us to observe dynamic scattering signal variations from anions and cations, which arise from their distinct optical polarizability and hence effective refractive indices~\cite{utterback2023operando}. 

The observed scattering intensity from each electrode in Figure~\ref{fgr:imaging_field}-b and c exhibit a periodic out-of-phase relationship with each other, aligning in phase with the applied potential, which is also visible in Figure~\ref{fgr:imaging_field}-d and e with the applied potential depicted in Figure~\ref{fgr:imaging_field}-f. Figure~\ref{fgr:imaging_field}-g displays the measured current passing between the electrodes. The exponential decay observed in each cycle is indicative of the EDL capacitive charging. A very small residual current will be present at longer waiting times, corresponding to Faradaic electrochemical reactions. For the experiments described here, residual Faradaic current can be neglected and we can assume that the electrode remains electrochemically inert throughout the measurement. In the supplementary materials, Figure S3 presents both Bode and cyclic voltammetry (CV) graphs, which provide detailed insights into the electrochemical properties of the system. More regions are selected from both electrodes to show that this integrated intensity change behavior is consistent in the shark-teeth electrodes, where ion current is out-of-phase in each electrode within the corresponding applied electric potential (Figure~S4). When repeating the experiment with identical parameters but employing a triangular waveform potential instead of a square one, we observe periodic fluctuations in ion migration intensity, aligning with the profile of the applied potential (Figure~S5). 

\subsection{Nanoparticle tracking with shark-teeth electrode }

\begin{figure}[!htbp] 
\centering
 \includegraphics[width=14 cm]{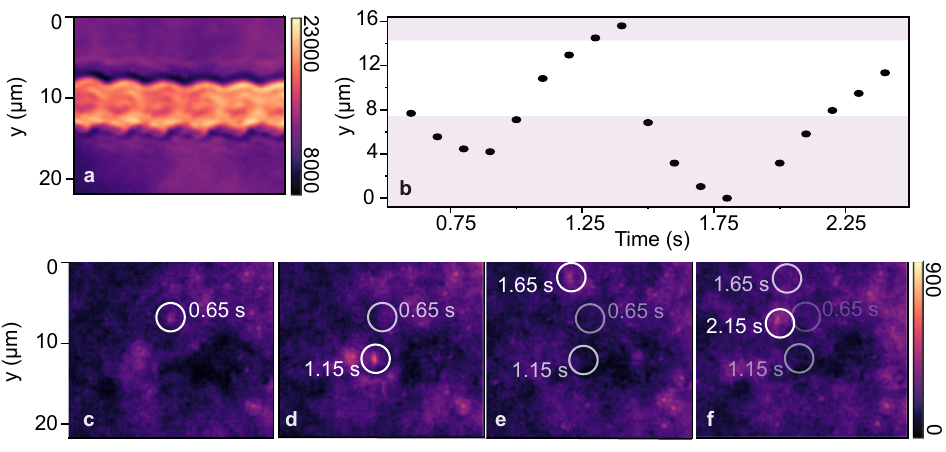}
  \caption{Particle tracing with shark-teeth electrodes under a 1 Hz square wave potential application: (a) Fluorescent microscopy image showing the shark-teeth electrode and particles. (b) Projected location of the nanoparticle over time, with the location of the shark-teeth electrodes indicated by the purple-shaded area. Particles are shown at (c) 0.65 s, (d) 1.15 s, (e) 1.65 s, and (f) 2.15 s, with the same particle marked with a white ring and previous locations fading in opacity in each subsequent frame. Figure (a) is extracted as a background. }
  \label{fgr:track}
\end{figure}

In the nanoparticle trace experiments described in this section, the electrolyte is replaced with an aqueous solution containing 200~nm fluorescent polystyrene particles. This experiment is a necessary first step toward potentiodynamic microscopy around trapped colloidal nanoparticles. Although smaller particles (down to 30 nm gold spheres in water) can be observed label-free by scattering via TIR microscopy (Supplementary Video~2), fluorescent microscopy is used exclusively for tracing the particles as particle tracking with a scattering microscope is prone to a much larger error due to continuously changing background speckle. The schematic of the fluorescence microscope is shown in Figure S6. The shark-teeth electrode is imaged via fluorescent microscopy and the result is shown in Figure~\ref{fgr:track}-a. Here, the darker regions are the gold electrodes and the brighter region corresponds to the gap in between. As we are not using a confocal microscope, the background fluorescence signal is much larger than each single particle. However, a single particle can be detected after subtracting the, on average, constant background. We have tracked the phoretic motion of a single nanoparticle under the application of a 1~Hz square-waved electrical potential, which is plotted in Figure~\ref{fgr:track}-b. The particles cyclically approach and recede from one electrode. This particle oscillation is also shown in Supplementary Video~3. Some frames of that video are provided at 0.65 s (Figure~\ref{fgr:track}-c), 1.15 s (Figure~\ref{fgr:track}-d), 1.65 s (Figure~\ref{fgr:track}-e), and 2.15 s (Figure~\ref{fgr:track}-f), with previous locations fading in opacity in each subsequent frame. In these frames, shown every 0.5 seconds, Figure~\ref{fgr:track}-a serves as a background and is depicted in Supplementary Video~3. The transparency of the thin electrode enables visualization of the particle beyond the conductive layers.

This experiment shows that the nanoparticle is responding electrophoretically to the applied potential, but the force is not enough in this case to overcome the repulsive electrostatic forces between the negatively charged glass surface and the charge-stabilized colloidal particle. 
To reach our aim of trapping these nanoparticles at the electrode edges, we use high-frequency AC dielectrophoresis, which has previously been shown to work on other electrode geometries~\cite{dogruyuksel2024origami}.

\subsection{Trapping nanoparticles with shark-teeth electrodes}

To perform iontronic microscopy on colloidal nanoparticles, we here explore the possibility of trapping them with higher frequencies using the shark-teeth electrode design. Once the particle is trapped, it could be irreversible due to van der Waals forces, in which case one can stop the AC potential and continue with the low-frequency potentiodynamic measurements.
In case the van der Waals force is insufficient for capturing the particles, one can still combine the low-frequency component for EDL modulation with the high-frequency trapping. It is important to note that the Nernst-Planck-Maxwell equations governing the ion dynamics are nonlinear, and there could be mixing between low and high-frequency responses that have to be accounted for in interpreting the results.

\begin{figure}[!htbp] 
\centering
 \includegraphics[width=14 cm]{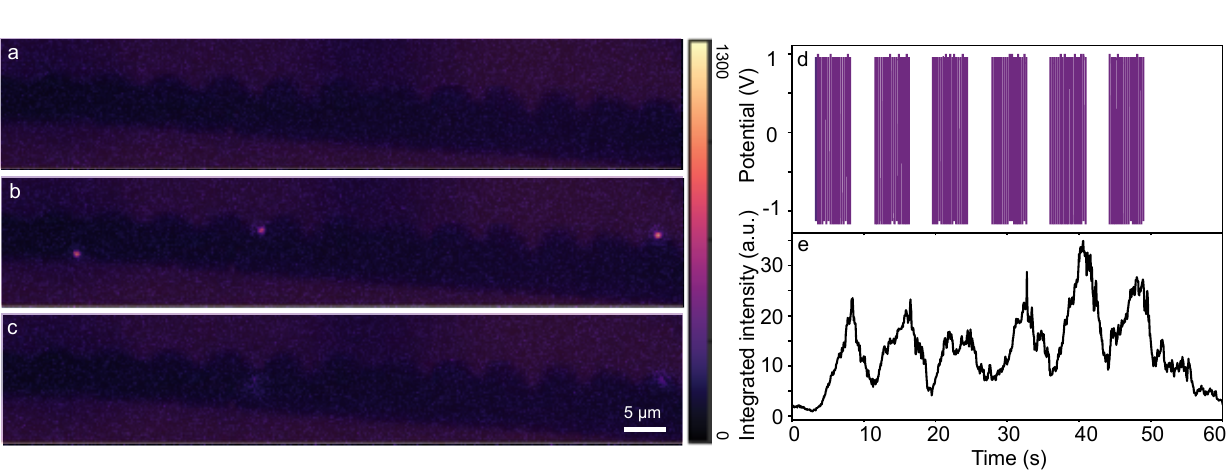}
  \caption{Particle trapping within shark-teeth electrode: (a) before, (b) during, and (c) after the application of an electric field at 100 kHz. (d) The applied potential is depicted where it is periodically terminated and initiated. (e) Integrated intensity changes during the applied potential on and off. Notably, upon application of the potential, three particles appear at the edges of the shark-teeth and vanish from the focus region when the potential is terminated.}
  \label{fgr:trapping}
\end{figure}

We applied a 100~kHz 2 V peak-to-peak square waveform potential to the electrodes while observing the nanoparticles on the fluorescence microscope. Particles are trapped at the sharp edges of the electrodes due to the gradient of the electric field but are released when the applied field is turned off. Figure~\ref{fgr:trapping}-a to c depict the process of trapping a few nanoparticles at the sharp edges, before applying the potential, during, and after turning off. At this frequency, the DEP effect is observable, where the gradient of the electric field exerts a force on the particles~\cite{zhang2019dep}. 
We can check the reversibility of this trapping effect by repeatedly turning the applied potential's on-and-off, as shown in Figure~\ref{fgr:trapping}-d. Correspondingly, Figure~\ref{fgr:trapping}-e illustrates the total change in the integrated fluorescence intensity around the electrode edge over time. From this measurement, most of the particles are released after each pulse, and the trapping mechanism is mostly reversible. Further details are available in Supplementary Video~4. Similarly, reversible particle trapping was observed with 30~nm gold nanoparticles using TIR microscopy at higher frequencies, allowing label-free visualization of the particles via scattering (see Figure S7 and Supplementary Video~2). 

We mention in passing that occasionally, at an intermediate frequency of around 100 Hz, particles are absorbed strongly enough to the electrodes and sometimes trapped irreversibly. 

In these intermediate frequency regimes, rapid oscillations prevent the stable formation of the EDL and hence the charged colloidal gets a higher chance of stochasticly passing through the Debye layer and getting trapped by van der Waals forces.

\section{Conclusion}

In conclusion, this study presents a novel method for fabricating laser-drilled gold electrodes with sharp edges (shark-teeth patterns) that are suitable for trapping nanoparticles and at the same time allow for detecting the ion dynamics with EDL-modulation microscopy. This facile electrode fabrication can be applied to most thin film conducting layers including those that are not easily compatible with clean-room lithographic methods, e.g. organic conducting layers or molecular assemblies.

At low frequencies, both ions and nanoparticles oscillate primarily due to EDL modulation. TIR microscopy confirmed this behavior, showing real-time intensity changes synchronized with the applied potential. The particle can also be used as a tracer to simultaneously detect the fluid motion.
As higher, dielectrophoresis (DEP) forces allow for the reversible trapping of nanoparticles. We demonstrated reversible DEP-based trapping at the sharp edges of the electrodes. This transition showcases the versatility of our electrode design in controlling different electrokinetic phenomena, allowing for the manipulation of solvated species at specific positions for further analysis.

This work presents a robust and scalable method for fabricating electrodes without lithography. Therefore, we can envision using this fabrication technique on substrates that are not compatible with lithography. Some examples are the tip of an optical fiber or a cantilever for atomic force microscopy. Further applications of this fabrication method will be investigated in future work.

\section{Materials and Methods}
\subsection{Physical vapor deposition}
A \#1.5 glass slide, previously plasma-cleaned, was coated with 1 nm of chromium and 10 nm of gold using the Edwards Coating System E306A through an H-shaped mask. The horizontal line of the H-mask has a width of 750 µm.

\subsection{Laser ablation}

The shark-teeth pattern is made using the setup described in \cite{vreugdenhil2022setup}. Although this setup is mainly used for laser ablation experiments\cite{vreugdenhil2023pulse} it has also proven useful for laser patterning\cite{cruciani2024patterning}. Figure S2 shows a simplified schematic overview of the setup. An ultrafast laser system produces laser pulses at a central wavelength of 1034 nm and a pulse duration of 170 ps. A combination of neutral density filters, a polarizing beamsplitter, and a half-wave plate control the pulse energy. For making the shark teeth electrodes, we set the pulse energy on the sample to approximately 28~nJ, which is equivalent to a fluence of 0.42 J/cm$^2$.

After the power control section, the pulses pass two pellicle beamsplitters. A microscope objective then focuses the pulses onto the gold surface. The light produced by a green LED is coupled into the beam path using the first of the two pellicle beamsplitters. This light illuminates the sample such that the sample can be monitored using the objective, the second pellicle beamsplitter, and a CCD camera. The sample is mounted on a motorized stage.

To make the shark teeth electrodes, we start with the plain H-sample, see Figure \ref{fgr:electrode_fabrication}-a. After leveling the sample, we run the laser at an effective repetition frequency of 2~kHz. The start position of the sample is such that the optical axis of the objective is located a short distance ($\approx$ 50~$\mu$m) from the inner gold strip of the sample and such that the surface of the sample is in the focal plane of the objective. We then open the shutter and start moving the stage at a speed of 10~m/s, such that the full width of the gold strip passes through the focus of the objective. We have chosen this combination of repetition frequency, pulse energy, and stage speed such that the disk-shaped ablation areas slightly overlap (see Figure \ref{fgr:electrode_fabrication}-b to f). When the full gold strip has passed through the focus of the objective, the shutter closes again. As a final step, we visually check the electrode for possible short circuits using the CCD camera. If there are any short circuits, we manually ablate the left-over gold that causes the short circuit by opening the shutter for a short moment.

\subsection{TIR microscopy}
A 640 nm laser beam (Ignis, Laser Quantum) with a power of 60 mW was focused on the back focal plane of an oil-immersion objective (Nikon CFI Apochromat TIRF 60×, 1.49 NA) by a tube lens with a focus lens of 200 mm. The focused beam was off-axis so that the beam coming out from the objective could illuminate the sample at a large angle, which enables the total internal reflection to occur at the electrode-electrolyte interface. The total internal reflected light was guided to a position detector (PQD-80A, Thorlabs) to maintain the position of the sample at the focus plane of the objective. The scattered light from the shark-teeth electrode is collected by the same objective and then focused by a tube lens to project it to a scientific CMOS (sCMOS) camera (Hamamatsu ORCA-Flash 4.0 V3).

\subsection{Fluorescent microscopy}
A blue LED (Thorlabs, M395L4) excites the particles, with the excitation light reflected by a dichroic filter towards the microscope objective (ZEISS 40×/0.65) via a mirror. The emitted light traveled back through a long-pass dichroic filter (500 nm) to the camera (Hamamatsu Digital Camera, C11440). This setup efficiently filtered out the source light, allowing clear particle visualization. 

\subsection{Electric Field Generation}
A waveform generator (KEYSIGHT InfiniiVision DSOX2024A) synchronized with the camera to monitor the applied voltage and circuit, enabling concurrent observation of the particles' response to the applied potential.
\subsection{Electrolyte solution}
Sigma-Aldrich potassium chloride (anhydrous, free-flowing, Redi-Dri™, ACS reagent, $ \geq $99\%) is mixed with water to prepare a 0.1 M electrolyte solution.

\subsection{Particle solutions}

Aqueous suspensions of Fluoresbrite® YG carboxylate microspheres (200 nm in diameter, 2.5\% weight/volume), with an excitation peak at 441 nm and an emission peak at 486 nm, were used for particle tracing and trapping experiments via fluorescence microscopy. For trapping experiments using a TIR microscope, Sigma-Aldrich 30 nm gold nanoparticles (OD 1) stabilized in citrate buffer were utilized.

\subsection{SEM}
Nanoelectrode morphology was analyzed using a ZEISS EVO 15 SEM operated at 5 kV. 

\subsection{Impedance Spectroscopy}
Impedance and cyclic voltammetry measurements of the shark-teeth electrodes were conducted using a BioLogic SP300 potentiostat.

\begin{acknowledgement}

The authors thank Kevin Namink, Kerman José Gallego Lizarribar, and Sjoerd Quaak for their assistance with some of the earlier measurements for this project. The authors also acknowledge technical support with the measurement setup from Paul Jurrius, Aron Opheij, Dante Killian, Jan Bonne Aan, and Arjan Driessen. This research was supported by Refeyn LTD and Nederlandse Organisatie voor Wetenschappelijk Onderzoek (Vici 68047618).

\end{acknowledgement}

\begin{suppinfo}

The Supporting Information is available. 

\end{suppinfo}

\bibliography{achemso-demo}

\end{document}










\renewcommand{\thefigure}{S\arabic{figure}}
\section{}
\subsection{Shark-teeth electrode }
\begin{figure}[h] 
\centering
 \includegraphics[width=16 cm]{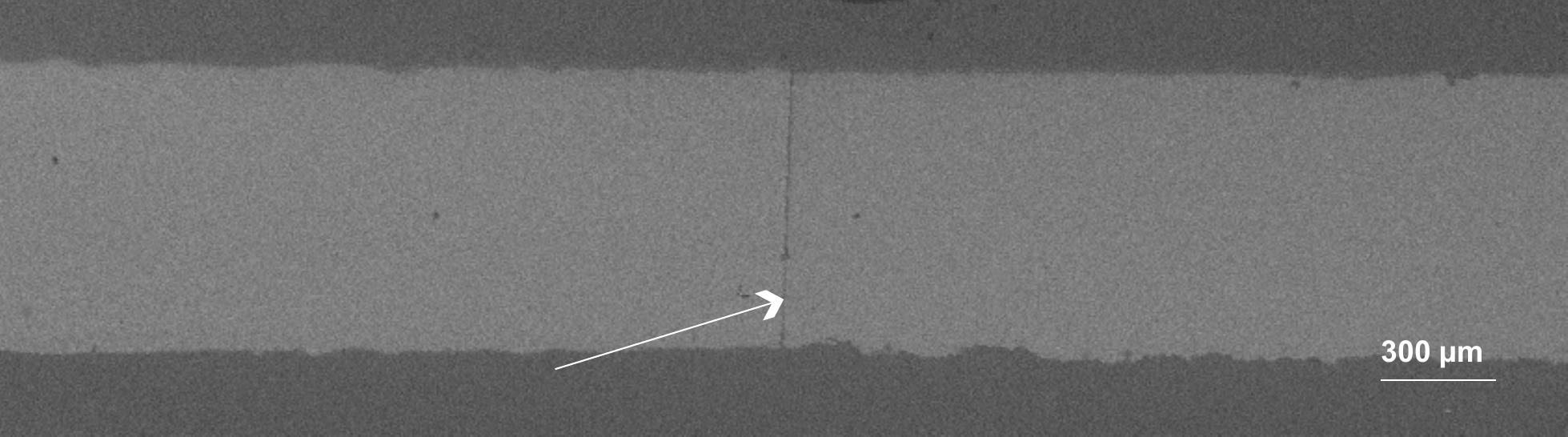}
  \caption{SEM of the laser ablated region on conductive layer coated on glass. }
  \label{fgr:SEMofSI}
\end{figure}

Figure~\ref{fgr:SEMofSI} displays scanning electron microscopy (SEM) image the laser-ablated region of the conductive coatings, which forms two planar electrodes.

\subsection{Laser ablation setup}
\begin{figure}[!htbp] 
\centering
 \includegraphics[width=0.6\linewidth]{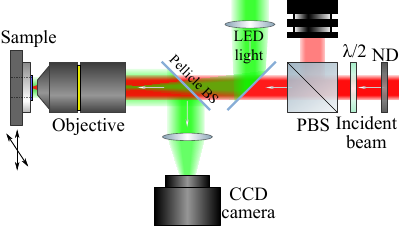}
  \caption{The simplified schematic of laser ablation setup.}
  \label{fgr:ablation_setup}
\end{figure}

The simplified schematic of laser ablation setup to make shark-teeth pattern is shown in Figure~\ref{fgr:ablation_setup}.

\subsection{Impedance spectroscopy }
\begin{figure}[h] 
\centering
 \includegraphics[width=16 cm]{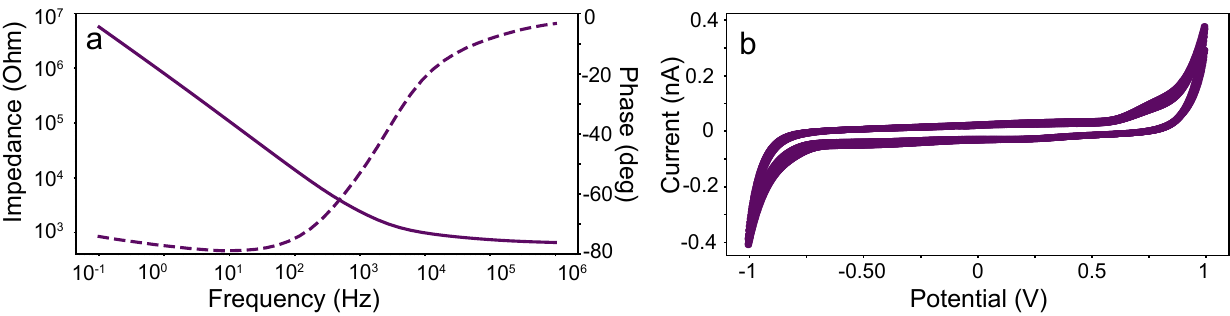}
  \caption{Impedance spectroscopy of shark-teeth electrode: (a) Bode plot of the impedance of the shark-teeth electrode with 0.1 M KCl for a sinusoidal 10 mV alternating potential. (b) Cyclic voltammetry of the shark-teeth electrode in contact with the electrolyte measured with 3 cycles. }
  \label{fgr:impedance}
\end{figure}

We further examined the shark-teeth electrode using 0.1 M KCl electrolyte through electrochemical impedance spectroscopy (EIS). The impedance spectrum of the shark-teeth electrodes measured at 10 mV (Figure~\ref{fgr:impedance}-a) reveals that, as the frequency increases, the magnitude of the impedance decreases (solid lines). At low frequencies, the magnitude of the impedance declines, and beyond 10 kHz, the variation in magnitude diminishes, indicating electrochemical stability in that frequency range. Moreover, the phase at higher frequencies approaches 0 (dashed lines). The cyclic voltammetry of the shark-teeth electrode, recorded across three cycles from -1 V to 1 V (Figure~\ref{fgr:impedance}-b), further confirms the electrochemical stability of the electrodes, with the CV curve showing consistency across the cycles.

\subsection{Ion accumulation variations at different regions }

\begin{figure}[h] 
\centering
 \includegraphics[width=16 cm]{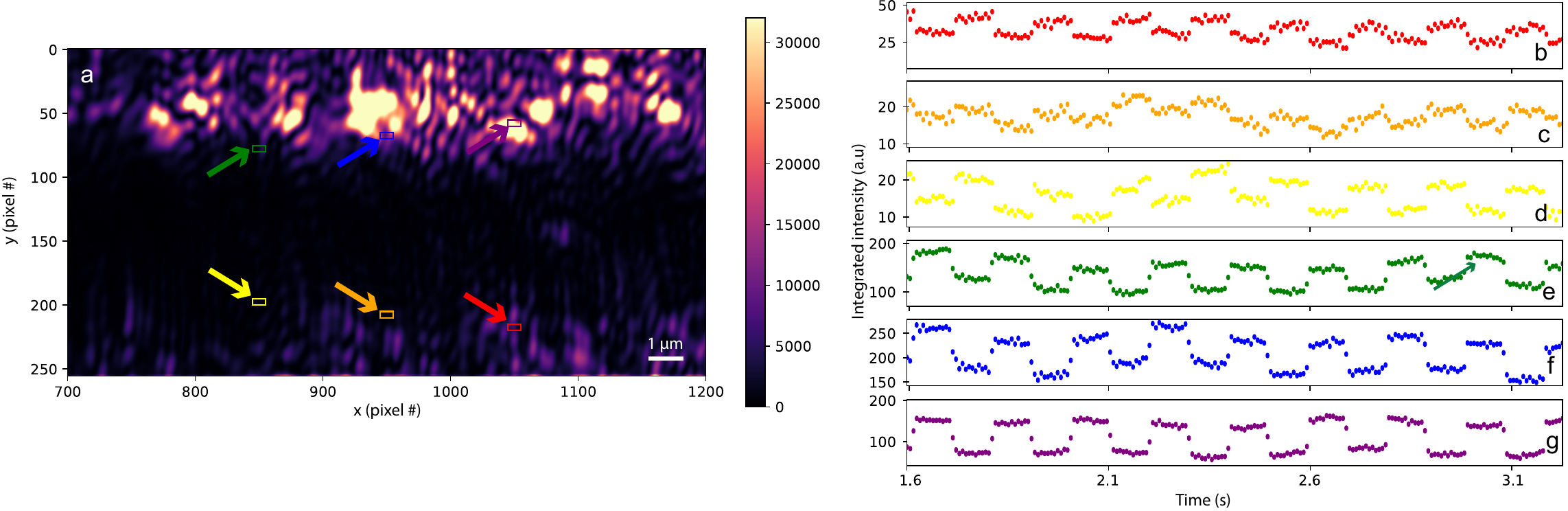}
  \caption{(a) Image of the shark-teeth electrode under a 5 Hz, 1 V peak-to-peak square waveform, showing six distinct regions marked with different colored squares and arrows. The integrated intensity extracted from the marked regions in (b) red, (c) orange, (d) yellow, (e) green, (f) blue, and (g) purple. }
  \label{fgr:different}
\end{figure}

In this section, we selected additional regions on each of the shark-teeth electrodes. These regions from the same electrode exhibit in-phase integrated intensity behavior over time, while those from opposing electrodes demonstrate out-of-phase behavior with each other and in-phase behavior within the same electrode (Figure~\ref{fgr:different}). 

\subsection{Ion accumulation variations at triangular waveform}
\begin{figure}[h] 
\centering
 \includegraphics[width=16 cm]{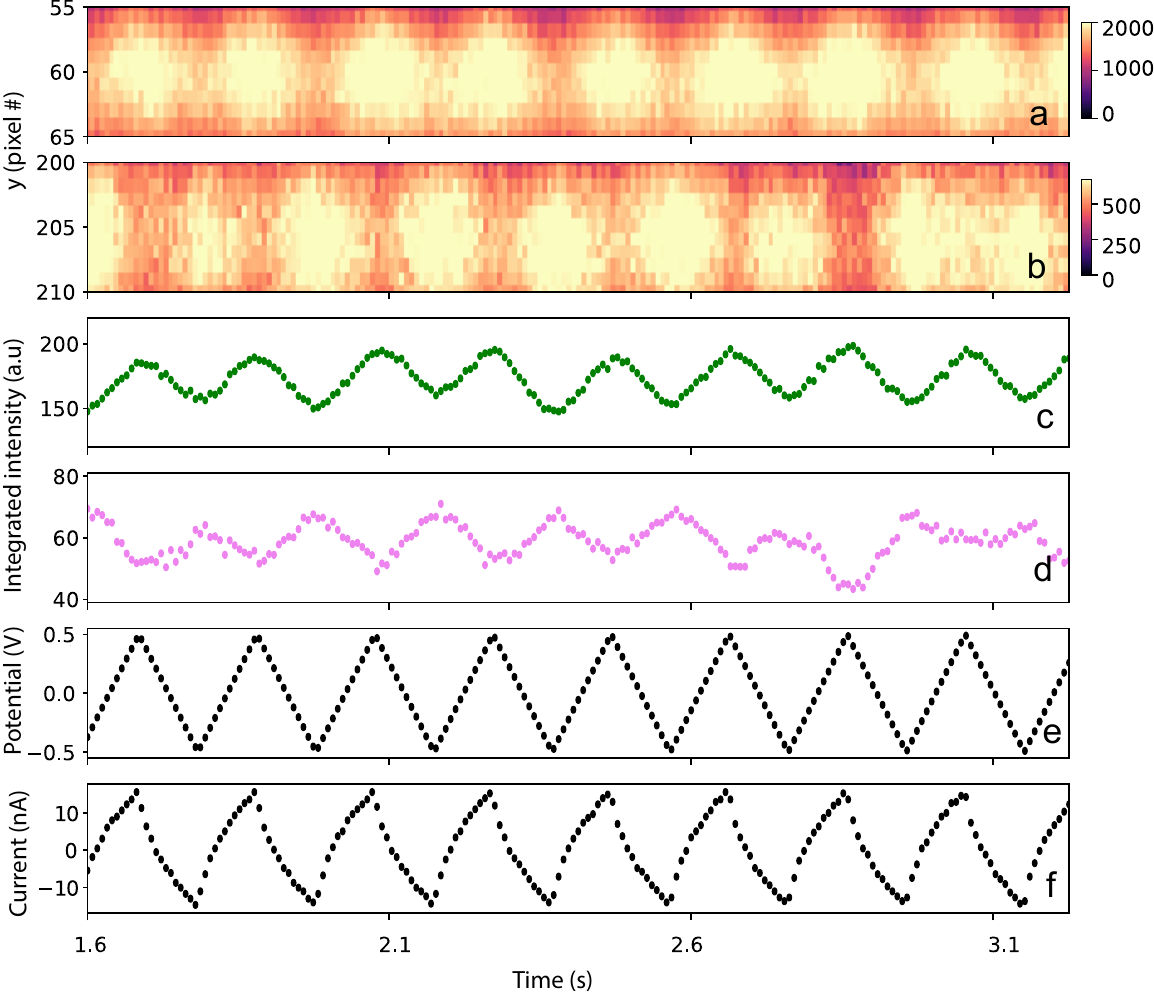}
  \caption{Chemographs of the shark-teeth electrodes: (a) top and (b) bottom electrode regions as depicted in Figure 2-a, under the influence of a 1 Hz, 1 V peak-to-peak triangular waveform. Integrated intensities extracted from (c) the top and (d) the bottom electrodes. (e) Applied electric potential and (f) current.}
  \label{fgr:triangle}
\end{figure}

In this section, we applied a triangular waveform to the electrode submerged in a KCl electrolyte. The resulting intensity variations are depicted in chemographs from opposite electrodes (see Figure~\ref{fgr:triangle}-a and b). Figure~\ref{fgr:triangle}-c and d show the integrated intensities from these electrodes. The regions used for data extraction are outlined in Figure 2-a of the main text.Additionally, Figure~\ref{fgr:triangle}-e and f display the applied potential and current, respectively. These figures illustrate an out-of-phase relationship between the opposing electrodes, emphasizing the intensity fluctuations within the applied field.

\subsection{Fluorescence Microscopy }

\begin{figure}[!htbp] 
\centering
 \includegraphics[width=16 cm]{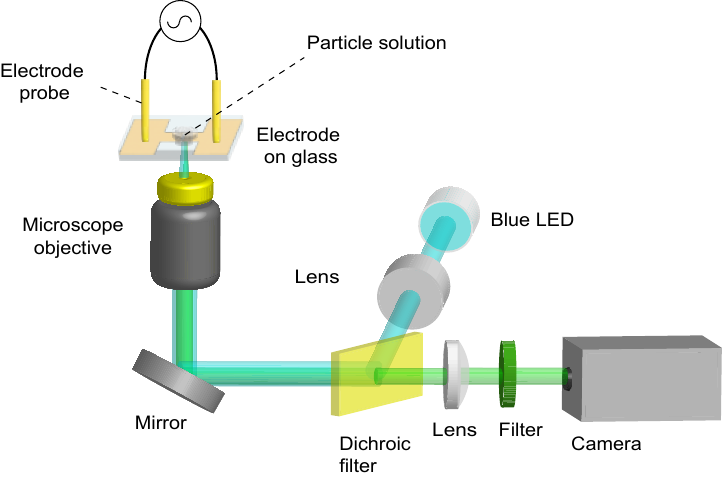}
  \caption{Schematic of the fluorescence microscopy used in fluorescent particle trace and trap experiments via shark-teeth electrode. }
  \label{fgr:fluorescent}
\end{figure}

The schematic of the custom-built fluorescence Microscopy is shown in Figure~\ref{fgr:fluorescent}.

\subsection{Label free imaging of 30 nm gold nanoparticle reversible trapping }

\begin{figure}[h] 
\centering
 \includegraphics[width=10 cm]{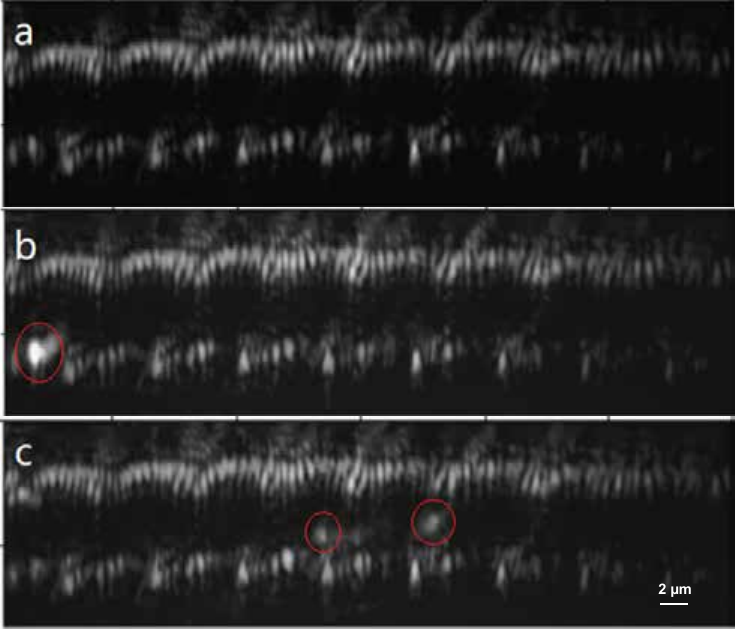}
  \caption{Label-free imaging of 30 nm gold nanoparticles during trapping: (a) Before, (b) during and (c) after the application of electric field, with detected particles marked with red circles.}
  \label{fgr:30nm}
\end{figure}

In Figure~\ref{fgr:30nm}, the scattering-based shark teeth electrode is depicted through TIR microscopy, with the addition of 30 nm gold nanoparticles at 2.56 V peak to peak  660 kHz potential application. Figure~\ref{fgr:30nm}-a illustrates the absence of particles prior to the application of an electric field. In Figure~\ref{fgr:30nm}-b, a trapped particle is highlighted by a red circle, indicating the efficacy of shark teeth electrodes in capturing smaller nanoparticles. Upon discontinuation of the electric field in Figure~\ref{fgr:30nm}-c, the particles are released, with specific ones being identified by red circles once more. This observation underscores the dynamics of label-free nanoparticle trapping capability of shark teeth electrodes observed via TIR microscopy.

